\documentclass[amsmath,amssymb,11pt,superscriptaddress,reprint, preprintnumbers, notitlepage,aps,twocolumn,nofootinbib]{revtex4-1}
\pdfoutput=1 
\usepackage[utf8]{inputenc}
\usepackage[english]{babel}
\usepackage{amsmath}
\usepackage{graphicx}
\usepackage{dcolumn}
\usepackage{pbox}
\usepackage{amssymb}
\usepackage{epsfig}
\usepackage[dvipsnames]{xcolor}
\usepackage[normalem]{ulem}
\usepackage{slashed}
\usepackage{amssymb}
\usepackage{ mathrsfs }
\usepackage{color}
\usepackage[font=small]{caption}
\usepackage[font=small]{subcaption}
\usepackage{url}
\usepackage[makeroom]{cancel}
\usepackage{tikz}
\usetikzlibrary{arrows}
\usepackage{enumitem} 
\usepackage{soul}
\usepackage{dsfont}
\usepackage{tabularx}

\definecolor{MyDarkBlue}{rgb}{0.1, 0.1, 0.8} 
\definecolor{SBlue}{rgb}{0.2, 0.4, 0.7} 
\definecolor{MyLightBlue}{rgb}{0.22,0.51,0.9}
\definecolor{MyGreen}{rgb}{0.0, 0.5, 0.0}
\definecolor{BrickRed}{rgb}{0.8, 0.25, 0.33}
\RequirePackage{hyperref}
\hypersetup{colorlinks, citecolor=SBlue,linkcolor=MyDarkBlue, urlcolor=PineGreen}

\definecolor{LightCyan}{rgb}{0.88, 1, 1}
\makeatletter
\renewcommand\@makecaption[2]{%
  \par
  \vskip\abovecaptionskip
  \begingroup
  
   \small\rmfamily
    \begingroup
     \samepage
     \flushing
     \let\footnote\@footnotemark@gobble
     \@make@capt@title{#1}{#2}\par
    \endgroup
  \endgroup
  \vskip\belowcaptionskip
}
\newcommand{\matrixel}[3]{\left< #1 \vphantom{#2#3} \right|
 #2 \left| #3 \vphantom{#1#2} \right>} 
\makeatother

\DeclareUnicodeCharacter{2212}{-}
\begin{document}

\allowdisplaybreaks
\title{\vspace{1cm}\Large 
Leptoquark-Mediated Two-Loop Neutrino Mass in Unified Theory
}

\author{\bf Kevin Hinze}
\email[E-mail:]{kevin.hinze@unibas.ch}

\author{\bf Shaikh Saad}
\email[E-mail:]{shaikh.saad@unibas.ch}

\affiliation{Department of Physics, University of Basel, Klingelbergstrasse\ 82, CH-4056 Basel, Switzerland}

\begin{abstract}
Scalar leptoquarks naturally arise within unified theories, offering a promising avenue for addressing one of the most significant challenges of the Standard Model--the existence of non-zero neutrino masses. In this work, we present a unified theory based on the SU(5) gauge group, where neutrino mass appears at the two-loop level via the propagation of scalar leptoquarks. Due to the unified framework, the charged fermion and neutrino masses and mixings are entangled and determined by a common set of Yukawa couplings. These exotic particles not only shed light on the neutrino mass generation mechanism but also help to achieve the unification of gauge couplings and are expected to lead to substantial lepton flavor violating rates, offering tangible opportunities for experimental verification. Reproducing the observed neutrino mass scale necessitates that a set of leptoquarks reside a few orders of magnitude below the unification scale--a specific feature of the proposed scenario. Moreover, maximizing the unification scale implies TeV scale new physics states, making them accessible at colliders. The diverse roles that leptoquarks play highlight the elegance and predictive ability of the proposed unified model. 
\end{abstract}

\maketitle
\section{Introduction}
So far, the Standard Model (SM) of particle physics has proven to be extremely successful in reproducing many experimental observations. Despite its triumphs, the SM predicts massless neutrinos, whereas neutrino oscillations have been experimentally confirmed. Therefore, the SM clearly requires an extension. In the quest to find an ultraviolet-complete theory beyond the SM, Grand Unified Theories~\cite{Pati:1973rp,Pati:1974yy, Georgi:1974sy, Georgi:1974yf, Georgi:1974my, Fritzsch:1974nn} (GUTs) have emerged as the most attractive scenarios. The SU(5) GUT, proposed in the early 1970s by Georgi and Glashow~\cite{Georgi:1974sy}, represents a significant step towards unifying the fundamental forces of nature. This unified framework has a number of appealing features, such as the merging of electromagnetic, weak, and strong nuclear forces into a single force at high energy levels; the unification of quarks with leptons; and the prediction of the existence of superheavy gauge bosons that mediate nucleon decays—a smoking-gun signature of GUTs.

However, the Georgi-Glashow model's simplicity, characterized by the inclusion of only an adjoint Higgs, $24_H$, for GUT symmetry breaking and a fundamental Higgs, $5_H$, for electroweak symmetry breaking, renders it incompatible with experimental findings. Specifically, the model falls short in several key areas: (i) it fails to achieve unification of gauge couplings, (ii) it results in a mass degeneracy between down-type quarks and charged leptons, which is inconsistent with observed data, and (iii) it does not provide a mechanism for incorporating neutrino masses.

Within the renormalizable scheme, one of the simplest options to restore gauge coupling unification, saving the theory from too rapid proton decay, as well as correcting the bad mass relations between the charged leptons and down-type quarks can be achieved by extending the Georgi-Glashow model by adding just a single Higgs in the $45_H$ dimensional representation. In this framework, neutrinos, however remain massless. There have been numerous efforts to endow neutrinos with non-zero masses (see, for example, Refs.~\cite{Dorsner:2005fq,Dorsner:2005ii,Bajc:2006ia,Dorsner:2006hw,Dorsner:2007fy,Antusch:2021yqe,Antusch:2022afk,Calibbi:2022wko,Antusch:2023kli,Antusch:2023mqe}, where neutrinos get their masses at the tree-level, and  Refs.~\cite{Wolfenstein:1980sf,Barbieri:1981yw,Perez:2016qbo,Kumericki:2017sfc,Saad:2019vjo,Dorsner:2019vgf,Dorsner:2021qwg,Antusch:2023jok,Dorsner:2024jiy,Klein:2019jgb}, where neutrino masses appear at the loop level\footnote{For original radiative neutrino mass models, see Refs. \cite{Cheng:1977ir, Zee:1980ai, Cheng:1980qt, Babu:1988ki}.}).

By observing the crucial fact that $5_H$ and $45_H$ multiplets already contain a number of scalar leptoquarks, in this work, we propose a two-loop neutrino mass mechanism by an economical extension of this framework. More specifically, we add scalar fields in the $40_H$-dimensional representation, submultiplets of which, following electroweak symmetry breaking, mix with leptoquarks contained within $5_H$ and $45_H$. This mixing facilitates interactions between leptons and quarks for some of the submultiplets of $40_H$, allowing neutrinos to acquire mass via quantum corrections. Intriguingly, in the last several years, leptoquarks have gained a lot of attention in the particle physics community due to their ability to address multiple flavor anomalies that include the longstanding tension in the muon anomalous magnetic moment.

It is interesting to point out that within the proposed setup, the masses of the charged fermions and neutrinos are intertwined and governed by a common set of original Yukawa interactions. We find that to correctly reproduce the neutrino mass scale, a set of leptoquarks running inside the loop must have masses several orders of magnitude below the GUT scale. However, in general,  some of the leptoquarks participating in neutrino mass generation facilitate proton decay, which typically imposes a lower limit of $M_\mathrm{LQ}\gtrsim 10^{12}$ GeV. Additionally, the Yukawa interactions tailored to fit the masses and mixings of charged fermions and neutrinos predict the occurrence of charged lepton flavor violating (cLFV) processes. These processes offer distinct avenues for experimental investigation. Current searches for cLFV significantly constrain the mass of leptoquarks involved in these processes to be above $M_\mathrm{LQ}\gtrsim 10^5$ GeV.  Therefore, in consistently reproducing the charged fermion mass spectrum and neutrino oscillation data, while meeting the requirements of experimental proton decay and cLFV bounds, two categories of leptoquarks emerge in this model: one group residing in the range of $10^{5-6}$ GeV, and a second group living close to $10^{12}$ GeV. Interestingly, to maximize the GUT scale, a set of new physics states that includes an iso-triplet, a color sextet, and a color octet scalar could potentially have masses close to the TeV range. This makes them accessible to exploration at present and forthcoming collider experiments, presenting an exciting frontier for experimental physics.

This article is organized in the following way. In Sec.~\ref{model}, we introduce the proposed model and derive the charged and neutral fermion masses and mixings matrices. In Sec.~\ref{results}, we describe gauge coupling unification, proton decay, and lepton flavor violation constraints. Moreover, a detailed numerical analysis taking into account fits to the charged fermion masses and mixings and neutrino oscillation data is also presented in Sec.~\ref{results}. Finally, we conclude in Sec.~\ref{conclusion}.

\begin{figure*}[th!]
\centering
\includegraphics[width=0.9\textwidth]{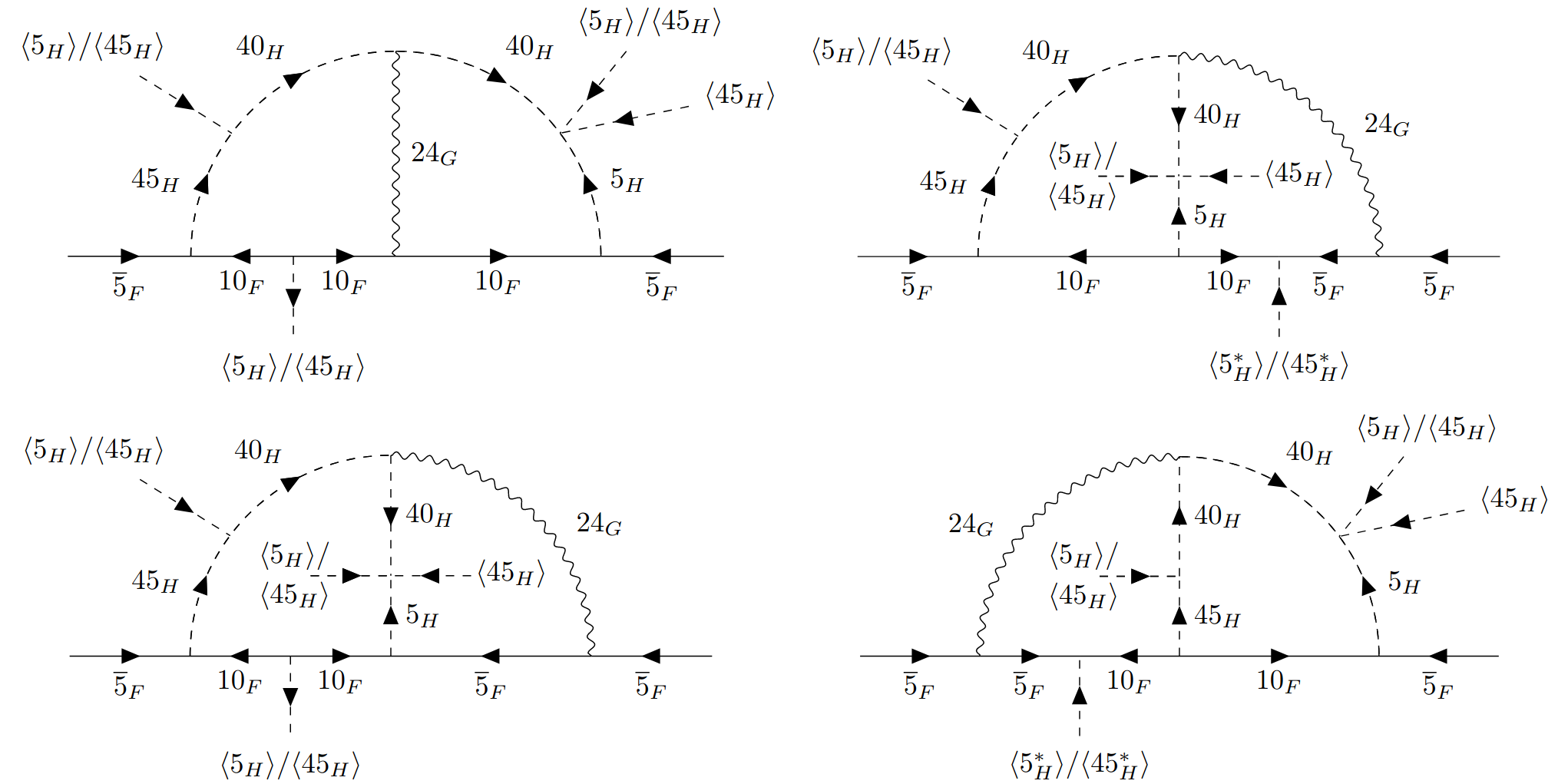}
\caption{Example Feynman diagrams generating  Majorana neutrino masses. See text for details. }
\label{fig:diagram}
\end{figure*}

\section{Model}\label{model}
In this section, we delve into the intricacies of the proposed model, specifically crafted to explain the phenomenon of neutrino oscillation. First, we describe how the bad mass relations between the down-type quarks and charged leptons are corrected within this framework. Subsequently, we provide a comprehensive exploration of the mechanisms underlying neutrino mass generation. 

In an SU(5) GUT, all the SM fermions of each generation are contained in the following two fermionic representations:
\begin{align}
&{\bm{\overline{5}_F}}_i= \ell_i (1,2,-\frac{1}{2}) \oplus d_i^c (\overline{3},1,\frac{1}{3}),  \\
&{\bm{10_F}}_i= q_i (3,2,\frac{1}{6}) \oplus u_i^c (\overline{3},1,-\frac{2}{3}) \oplus e_i^c (1,1,1),  
\end{align} 
where $i=1-3$ is the generation index. In this scheme, the GUT symmetry is first broken down to the SM gauge group by the adjoint representation, $24_H$, and finally, the electroweak symmetry breaking is assisted by $5_H+45_H$ Higgses: 
\begin{align}
SU(5) &\xrightarrow[]{\langle 24_H\rangle} SU(3)_C \times SU(2)_L \times U(1)_Y 
\\
&\xrightarrow[]{\langle 5_H\rangle, \langle 45_H\rangle} SU(3)_C \times U(1)_\mathrm{em}  \;.  
\end{align}
We define these scalar fields in the following way:
\begin{align}
\bm{5_H} \equiv \phi &= \phi_1 (1,2,\frac{1}{2}) \oplus \phi_2 (3,1,-\frac{1}{3}), \\
\bm{24_H} \equiv \Phi  &= \Phi_1 (1, 1, 0) \oplus \Phi_2 (1, 3, 0)\oplus \Phi_3 (8, 1, 0)
\nonumber\\&
\oplus \Phi_4 (3, 2, -\frac{5}{6} ) \oplus \Phi_5 (\overline{3}, 2, +\frac{5}{6} ),  \\
\bm{45_H} \equiv \Sigma &= \Sigma_1 (1,2,\frac{1}{2}) \oplus \Sigma_2 (3,1,-\frac{1}{3}) \oplus \Sigma_3 (\overline{3},1,\frac{4}{3})
\nonumber\\&
\oplus \Sigma_4 (\overline{3},2,-\frac{7}{6}) \oplus \Sigma_5 (3,3,-\frac{1}{3}) 
\nonumber \\ & 
\oplus \Sigma_6 (\overline{6},1,-\frac{1}{3}) \oplus \Sigma_7 (8,2,\frac{1}{2}).  
\end{align}  
Furthermore, the electroweak scale vacuum expectation values (vevs) are defined as $\langle 5_H\rangle = v_{5}/\sqrt{2}\left(0\;0\;0\;0\;1\right)^T$ and $\langle 45_H\rangle_{1}^{15}=\langle 45_H\rangle_{2}^{25}=\langle 45_H\rangle_{3}^{35}=-1/3\langle 45_H\rangle_{4}^{45}=v_{45}/(2\sqrt{6})$, with $\sqrt{v_{5}^2+v_{45}^2}=246$ GeV.

\subsection{Charged fermion mass}
With the fields given above, the complete Yukawa Lagrangian is given by~\cite{Georgi:1979df}
\begin{align}
-\mathcal{L}_Y&=Y_A10_F\overline 5_F 5^*_H  + Y_B10_F10_F 5_H \nonumber\\
&+Y_C10_F\overline 5_F 45^*_H  + Y_D10_F10_F 45_H\;. \label{yukawa}
\end{align}
Here, to avoid cluttering, we have suppressed the group as well as generation indices. Expanding the above Lagrangian and substituting electroweak scale vevs,  the fermion masses,  can be written as
\begin{align}
\mathcal{L} \supset -u^TM_Uu^c-d^TM_Dd^c-e^TM_Ee^c+\mathrm{h.c.}\,.
\end{align}  
From Eq.~\eqref{yukawa}, the charged fermion masses read~\cite{Georgi:1979df,Dorsner:2006dj}
\begin{align}\label{eq:MD}
&M_D=\frac{v_5}{2}Y_A-  \frac{v_{45}}{2\sqrt{6}}Y_C\;, \\\label{eq:ME}
&M_E=\frac{v_5}{2}Y_A^T+  \frac{\sqrt{3}v_{45}}{2\sqrt{2}}Y_C^T\;, \\\label{eq:MU}
&M_U=\sqrt{2}v_5 \left(Y_B+Y_B^T\right)+  \frac{v_{45}}{\sqrt{3}}\left(Y_D-Y_D^T\right)\;, 
\end{align}
where $D, E,$ and $U$ represent down-type quark, charged lepton, and up-type quark sectors, respectively. Note that due to the additional contribution from $45_H$, the bad mass relation of the Georgi-Glashow model $M_E=M^T_D$ is broken, and $M_U$ is no longer a symmetric matrix. We bi-diagonalize these matrices with the help of unitary matrices as follows:
\begin{align}
&M_U= U_L M_U^\mathrm{diag} U_R^\dagger\,,  \label{eq:massU}
\\
&M_D= D_L M_D^\mathrm{diag} D_R^\dagger\,,  \label{eq:massD}
\\
&M_E = E_L M_E^\mathrm{diag} E_R^\dagger\,. \label{eq:massE}
\end{align}
As we will show shortly, just like the charged fermions, the neutrino mass matrix is also determined by the same Yukawa couplings, namely $Y_A, Y_B, Y_C,$ and $Y_D$, making this scenario highly attractive and economical.

\subsection{Neutrino mass}
As aforementioned, the set of fields introduced in the preceding section cannot incorporate neutrino mass; further extension is needed. For that purpose, first observe that $5_H$ contains a scalar leptoquark $\phi_2 (3,1,-\frac{1}{3})$, whereas $45_H$ contains four scalar leptoquarks, namely $\Sigma_2 (3,1,-\frac{1}{3}), \Sigma_3 (\overline{3},1,\frac{4}{3}), \Sigma_4 (\overline{3},2,-\frac{7}{6})$, and $\Sigma_5 (3,3,-\frac{1}{3})$. With this inspection, in aiming to address non-zero neutrino masses, we introduce the following scalar field: 
\begin{align}\label{Higgs40}
\bm{40_H} \equiv \eta &= \eta_1 (1,2,-\frac{3}{2}) \oplus \eta_2 (\overline{3},1,-\frac{2}{3}) \oplus \eta_3 (3,2,\frac{1}{6}) 
\nonumber\\&
\oplus \eta_4 (\overline{3},3,-\frac{2}{3}) \oplus \eta_5 (\overline{6},2,\frac{1}{6}) \oplus \eta_6 (8,1,1).
\end{align}
Remarkably, $40_H$ contains submultiplets that share the same quantum numbers under the group $SU(3)_C \times U(1)_\mathrm{em}$ as in the leptoquarks residing in $5_H$ and $45_H$ Higgses. 
Consequently, the introduction of this field\footnote{Alternatively, one can employ a $35_H$-dimensional representation instead of the $40_H$-dimensional representation and obtain similar loop diagrams.  This is because $35_H\supset (\overline{3},3,-\frac{2}{3})$. If $35_H$ is employed, then in the neutrino mass diagram  of Fig.~\ref{fig:diagram}, the cubic scalar interaction arise from $45_H^2 35_H^*$ and the quartic scalar interaction emerge from $5_H^3 35_H$.} In this work, we stick to the $40_H$-dimensional representation. allows neutrinos to acquire masses at the two-loop order, as shown in Fig.~\ref{fig:diagram} (only a set of example diagrams are presented).

We emphasize that these Feynman diagrams are shown in the interaction basis. Neutrino mass must be computed in a basis where the fields running inside the loop are the physical states. Although in the flavor basis $\bm{40_H}$ has no Yukawa interactions, owing to electroweak symmetry breaking, its submultiplets $\eta_2 (\overline{3},1,-\frac{2}{3}), \eta_3 (3,2,\frac{1}{6})$, and $\eta_4 (\overline{3},3,-\frac{2}{3})$ can, in general, mix with the rest of the leptoquarks (contained within $5_H$ and $45_H$ Higgses) with the right quantum numbers and gain lepton-quark interactions--resulting in non-zero neutrino mass. These desired mixings originate (see Fig.~\ref{fig:diagram}) from cubic couplings of the form
\begin{align}
V\supset \mu_1 45_H 40^*_H 5_H + \mu_2 45_H 40^*_H 45_H \;,    
\end{align}
and quartic couplings of the type
\begin{align}
V&\supset \lambda_1 45_H 40_H 5_H 5_H + \lambda_2 45_H 40_H 5_H 45_H \nonumber\\&+ \lambda_3 45_H 40_H 45_H 45_H\;.    
\end{align}
As before, here, the group indices are suppressed. As can be seen from these two equations, mixings among leptoquarks are induced as a result of electroweak symmetry breaking.

Inside the loop, in addition to the $W^\pm$ gauge boson,  two types of leptoquarks propagate, LQ$^{\pm 1/3}$ and LQ$^{\pm 2/3}$. Note that with a single $40_H$ field, the LQ$^{\pm 1/3}$ (LQ$^{\pm 2/3}$) states are, in general, mixtures of $\phi_2, \Sigma_{2,5}, \eta_{3,4}$ ($\Sigma_{4,5}, \eta_{3,4}$) fields. Due to the presence of a large number of mixed leptoquarks, analyzing the entire parameter space is somewhat challenging. Therefore, we make a simplified assumption and focus only on the submultiplets $\phi_2$ within $5_H$, $\Sigma_4$ within $45_H$, and $\eta_4$ within $40_H$. In this scenario, physical leptoquarks LQ$^{\pm 2/3}$ (LQ$^{\pm 1/3}$)  are admixtures of $\Sigma_4^{\pm 2/3}$ ($\phi_2^{\pm 1/3}$) and  $\eta_4^{\pm 2/3}$ ($\eta_4^{\pm 1/3}$) states. LQ$^{\pm 2/3}$ states do not mediate proton decay, and as aforementioned, they must reside several orders of magnitudes below the GUT scale. Since both $\eta_4^{\pm 2/3}$ and $\eta_4^{\pm 1/3}$ are contained within the same multiplet, $\eta_4$, they exist around the same scale. Consequently, this can lead to problems with overly rapid proton decay primarily caused by states that are mostly composed of $\eta_4^{\pm 1/3}$ fields (the details of proton decay are discussed in Section~\ref{PD}). To circumvent this issue, we introduce a second copy of the $40_H$-dimensional representation, denoted as $\eta^\prime$. We further assume that the LQ$^{\pm 2/3}$ (LQ$^{\pm 1/3}$) states, which contribute to neutrino masses, are predominantly mixed states of $\Sigma_4^{\pm 2/3}$ ($\phi_2^{\pm 1/3}$) and $\eta_4^{\pm 2/3}$ ($\eta_4^{\prime\pm 1/3}$), with the remaining mixing angles set to very small values. This arrangement can be easily achieved by appropriately selecting the relevant coefficients of the cubic (quartic) terms in the scalar potential. In this way, we successfully decouple the masses of the lighter LQ$^{\pm 2/3}$ states from those of the heavier LQ$^{\pm 1/3}$ fields. For the physical leptoquarks that participate in the neutrino mass generation mechanism, we denote them by $\chi^{\pm Q}_a$, where $Q=1/3, 2/3$ and $a=1,2$. The leptoquarks LQ$^{\pm 5/3}$ residing in $\Sigma_4$ and $\eta^{(\prime)}_4$ do not contribute to neutrino mass nor do they mediate proton decay. However, they participate in rare lepton flavor violating processes. The remaining leptoquarks mentioned above are set to the GUT scale, and therefore, provide a completely negligible contribution to neutrino masses.

As mentioned in the introduction, due to the unified nature of our proposal, the neutrino mass matrix is not arbitrary and, therefore, does not decouple from the charged fermion masses and mixings.
By deriving the neutrino mass matrix, we find 
\begin{align}
M_N&= N M_N^\mathrm{diag} N^T
\\
&=-\frac{3g^2}{\sqrt{2}\left(16\pi^2\right)^2} \bigg\{
2Y^T_L M_U^\mathrm{diag} F_L\hat I + M_E^\mathrm{diag} Y^\dagger_R F_L \widetilde I
\nonumber\\&
+  M_E^\mathrm{diag} Y^T_L F_R^*  \overline I
\bigg\} +\mathrm{(transpose)}\;. \label{eq:nu-mass}
\end{align}
In the above equation, we have defined
\begin{align}
&Y_L=U^\dagger_L \left( \frac{-Y_A}{\sqrt{2}} \right)  E^*_L, \label{eq:nu-yuk-1}
\\
&Y_R=U^T_R \left( 2(Y^T_B+Y_B) \right)  E_R,
\\
&F_L=U^T_R \left( Y_C\right)  E^*_L,
\\
&F_R=U^\dagger_L \left( \sqrt{2}(-Y^T_D+Y_D) \right)  E_R\;. \label{eq:nu-yuk-4}
\end{align}
Since we compute the neutrino mass in the charged lepton mass diagonal basis, $N=U_\mathrm{PMNS}$ is the Pontecorvo–Maki–Nakagawa–Sakata (PMNS) matrix. 
The loop factors $\hat I, \widetilde I, \overline I$ appearing in Eq.~\eqref{eq:nu-mass} have complicated structures, and their full forms are given in Ref.~\cite{Julio:2022ton}. For our numerical analysis, although we utilized the exact loop functions, here we present a simplified form. In the limit when all the charged fermion masses can be neglected, one obtains $\hat I= \widetilde I= \overline I\equiv J_0$, which can be written in the following form~\cite{Julio:2022bue}:  \color{black}
\begin{align}
J_0 = \frac{1}{4}\sin2\theta\sin2\phi\sum_{a,b=1}^2(-1)^{a+b} J(M_{a+2},M_b). \label{eq4}
\end{align}
Here, $\theta$ and $\phi$ represent the mixing angles of leptoquarks with electric charges $1/3$ and $2/3$, respectively. Moreover, $M_{1,2}$ are masses for $2/3$ charged and $M_{3,4}$ are for $1/3$ charged leptoquarks. Finally, $J$ is defined in the following way:
\begin{align}
&J(M_{a+2},M_b) = 8\left(1-\frac{3}{4}\frac{M_{a+2}}{M_b}\right) - \frac{5\pi^2}{6} \left( 1-\frac{2}{5}\frac{M_{a+2}}{M_b} \right) \nonumber \\
& + \frac{M_b^2}{m_W^2} \left(\frac{M_{a+2}}{M_b} -1 \right)\left(2-\frac{\pi^2}{3}\right)   + \left(\frac{2M_{a+2}}{M_b}-1 \right)\ln \frac{m_W^2}{M_b^2},
\end{align}
which is valid in the regime of our interest, namely $M_b/M_{a+2} -1 \ll 1$. Here $m_W$ represents the mass of the $W^\pm$ gauge boson that runs inside the neutrino mass loop diagram Fig.~\ref{fig:diagram}.

\begin{figure*}[th!]
    \includegraphics[width=0.45\textwidth]{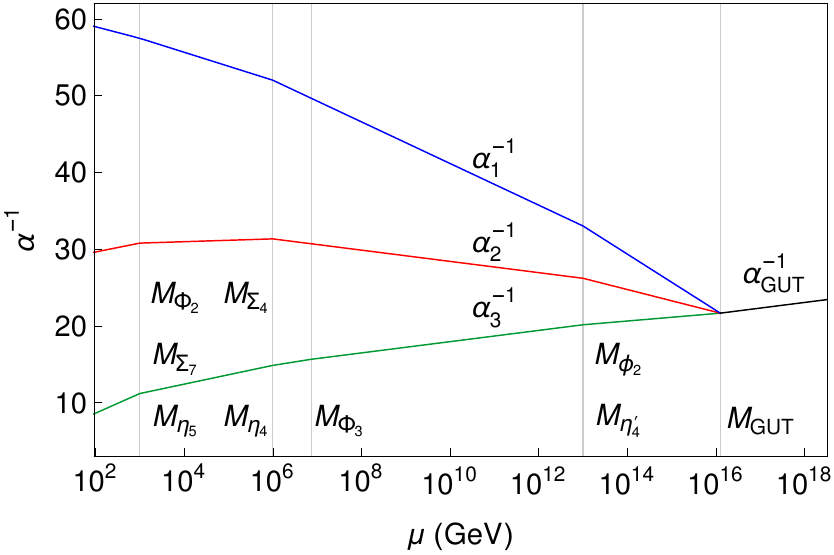}
    \hfill
    \includegraphics[width=0.48\textwidth]{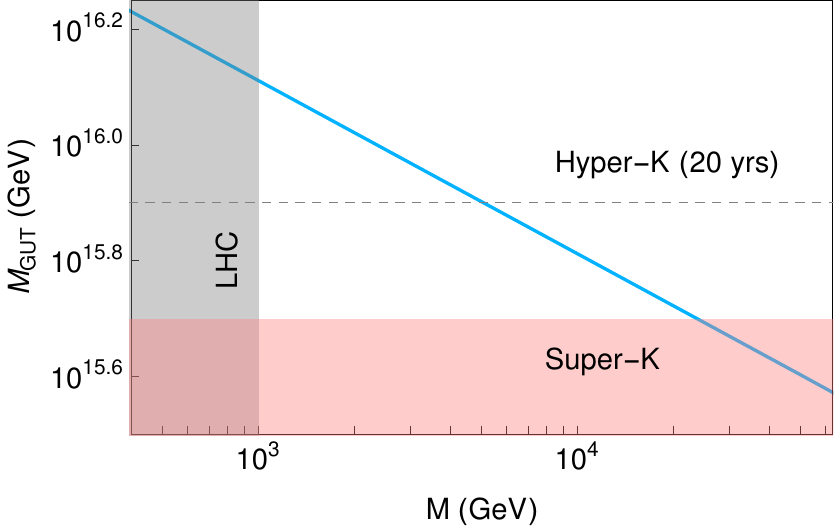}
    \caption{Left panel: Example of successful gauge coupling unification maximizing the GUT scale. Right panel: Maximal GUT scale as a function of the lightest intermediate-scale particle mass $M$. The lightest particle is either $\Phi_2$, $\Sigma_7$, or $\eta_5$. See text for details.}
    \label{fig:gcu}
\end{figure*}

\section{Results}\label{results}
In this section, we describe how gauge coupling unification is achieved and examine proton decay and lepton flavor violation appearing in the proposed model. After discussing these issues, we carry out a detailed numerical analysis, including a fit to the fermion masses and mixings.

\subsection{Gauge coupling unification}
The one-loop running of the SM gauge couplings ($\alpha_i=g_i^2/(4\pi)$) is given by 
\begin{flalign}\label{eq:gauge coupling running}
    \resizebox{0.87\linewidth}{!}{$\alpha_i^{-1}(M_Z)=\alpha_{\textrm{GUT}}^{-1}-\frac{b_i^{\textrm{SM}}\ln(\frac{M_\textrm{GUT}}{M_Z})}{2\pi}-\sum_J \frac{b_i^J\ln(\frac{M_{\textrm{GUT}}}{M_J}) }{2\pi}$},
\end{flalign}
where $b_i^{\textrm{SM}}$ are the SM one-loop gauge coefficients, while $b_i^J$ denote the one-loop gauge coefficients of intermediate-scale particles with masses $M_J$, such that $M_Z\leq M_J\leq M_{\textrm{GUT}}$. We take the experimental low energy values of the gauge couplings from Ref.~\cite{Antusch:2013jca}. Because of the neutrino mass generation mechanism we assume that the LQs $\Sigma_4,$ ($\phi_2$) and $\eta_4$ ($\eta_4^\prime$) have nearly degenerate masses $M_\mathrm{LQ}^\mathrm{light}$ ($M_\mathrm{LQ}^\mathrm{heavy}$). We then study a minimal scenario, where we freely vary the masses $M_{\Phi_2},M_{\Phi_3},M_{\Sigma_7},M_{\eta_5},M_{\mathrm{LQ}}^\mathrm{light}$ and $M_{\mathrm{LQ}}^\mathrm{heavy}$, and let the other states reside at the GUT scale. Then, the following relations between the intermediate-scale particle masses and the GUT scale can be derived from Eq.~\eqref{eq:gauge coupling running}:
\begin{align}\label{eq:gcu condtion 1}
    &M_{\mathrm{GUT}}=\sqrt[\leftroot{-1}\uproot{2}\scriptstyle 124]{\frac{M_Z^{371}M_{\phi_2}^{3}}{M_{\Phi_3} M_{\Sigma_7}^{28} M_{\eta_5}^{48}(M_\mathrm{LQ}^\mathrm{heavy})^{73}}}\;\times
    \nonumber\\
    &\hspace{2mm}\resizebox{0.85\linewidth}{!}{$\times\exp\left[\frac{\pi}{56}\left(25\alpha_1^{-1}(M_Z)-24\alpha_2^{-1}(M_Z)-\alpha_3^{-1}(M_Z)\right)\right]$},
    \\\label{eq:gcu condtion 2}
    &M_\mathrm{LQ}^\mathrm{light}=\sqrt[\leftroot{-1}\uproot{2}\scriptstyle 7]{\frac{M_Z M_{\phi_2}^{2}M_{\Phi_3}^{2}}{M_{\Phi_2}^3 M_{\eta_5}^2 M_{\eta_4^\prime}^{12} }}\;\times
    \nonumber\\
    &\hspace{2mm}\resizebox{0.85\linewidth}{!}{$\times\exp\left[\frac{2\pi}{7}\left(5\alpha_1^{-1}(M_Z)-9\alpha_2^{-1}(M_Z)+4\alpha_3^{-1}(M_Z)\right)\right]$}.
\end{align}
We freely vary the masses $M_{\Phi_2},M_{\Phi_3},M_{\Sigma_7},$ and $M_{\eta_5}$ between the TeV (for the compatibility of the LHC bounds, a lower limit of 1 TeV has been taken) and the GUT scale, while taking an upper bound of $10^7$~GeV on $M_{\mathrm{LQ}}^\mathrm{light}$ (which is needed to reproduce the correct neurino mass scale). Moreover, for $M_{\mathrm{LQ}}^\mathrm{heavy}$ we take a lower bound of $10^{12}$~GeV which is required due to proton decay constraints (cf.~subsequent subsection).
Thereby, we find that gauge coupling unification can easily be achieved. One example for gauge coupling unification is presented in the left panel of Fig.~\ref{fig:gcu}. Moreover, the right panel of Fig.~\ref{fig:gcu} shows how the maximal GUT scale depends on the smallest intermediate-scale particle mass $M$, where we have defined $M=\min\lbrace M_J\rbrace$. It turns out that in our study, the maximal GUT scale for a given $M$ is obtained for the scenario when $M=M_{\Phi_2}=M_{\Sigma_7}=M_{\eta_5}$. In particular, for $M=1$~TeV we find $M_{\textrm{GUT}}\leq1.5\times 10^{16}$~GeV. Also, gauge mediated proton decay roughly restricts the GUT scale to be above $M_{\textrm{GUT}}\gtrsim 5\times 10^{15}$~GeV. This, within the part of the parameter space we are exploring, gives a rough upper bound of $M\lesssim 25$~TeV. Hyper-Kamiokande--if it does not observe proton decay--will further reduce this scale to 5~TeV. If these states, namely the iso-triplet $\Phi_2(1,3,0)$, the color sextet $\Sigma_7(\overline 6, 2, 1/6)$ and color octet $\eta_5(8,2,1/2)$ reside close to the TeV scale, it opens up the possibility to search them at colliders.

\subsection{Proton decay}\label{PD}
Proton decay is mediated by the usual superheavy gauge bosons $X(3,2,-5/6)$ and $\overline{X}(\overline{3},2,5/6)$. Among the scalars, a subset of leptoquarks--specifically $\phi_2, \Sigma_2,$ and $\Sigma_5$--play a direct role in facilitating dimension six proton decay. Additionally, $\eta_3^{(\prime)}$ and $\eta_4^{(\prime)}$ contribute through their electroweak mixing with other leptoquarks, as elaborated in the preceding discussion. Furthermore, $\Sigma_3$ can also lead to proton decays via loop-level contributions. In this work, we have performed an in-depth analysis of proton decay rates at the tree level. A comprehensive list of two-body proton decay widths mediated by gauge and scalar bosons, relevant Wilson coefficients, as well as long- and short-distance coefficients  are presented in the Appendices \ref{sec:Apx-01} and \ref{sec:Apx-02}, respectively.  Moreover, the current experimental constraints and future sensitivities for two-body proton decay channels are listed in Table~\ref{tab:nucleon_decay}.

As mentioned in the previous section, with a single copy of $40_H$, keeping LQ$^{\pm 2/3}$ states light would also require keeping LQ$^{\pm 1/3}$ states light. However, the latter fields mediate proton decay. Instead of introducing a second copy of $40_H$, one might hope to entirely cancel all proton decay mediated by LQ$^{\pm 1/3}$ states. Therefore, realizing a light scalar leptoquark for neutrino mass generation necessitates a mechanism to suppress proton decay. The procedure to suppress proton decay using the freedom in the fermion mass matrices is widely known in the literature and has been studied, for example,  in Refs.~\cite{Nandi:1982ew,Berezinsky:1983va,Bajc:2002bv,FileviezPerez:2004hn,Dorsner:2004jj,Dorsner:2004xa,Dorsner:2005ii,Dorsner:2005fq,Nath:2006ut}  for the case of gauge boson mediated proton decay and in Refs.~\cite{Nath:2006ut,Dorsner:2012nq, Dorsner:2012uz, Fornal:2017xcj} for the case of scalar mediated proton decay. By adopting a similar approach, first note that the tree-level proton decay contributions of $\chi^{\pm 1/3}_a$ can be fully rotated away in our model.    In particular, rotating away this corresponding proton decay contributions  give the conditions~\cite{Dorsner:2012nq} $(U_L^\dagger(Y_B+Y_B^T)D_L^\ast)_{1\beta}=(D_R^\dagger Y_A^\dagger U_R^\ast)_{\beta1}=0$ for $\beta=1,2$; see Appendix~\ref{sec:Apx-02} for details.

However, imposing these constraints does not simultaneously suppress loop-level proton decay rates. Proton decay rates to two-body final states via one-loop diagrams, for example, through scalar leptoquarks and $W$-bosons in the loop, are derived and summarized in Appendix~\ref{sec:Apx-02}. In fact, cancelling the tree-level proton decays enhances the loop-induced processes that leads to $M_\mathrm{LQ}\gtrsim 10^{13}$ GeV. For consistency in reproducing the correct neutrino mass scale, replicating the observed charged fermion mass spectrum, and evading current stringent proton decay bounds, we do not utilize a cancellation mechanism. Instead, we introduce a second copy of $40_H$, as elaborated in the preceding section. \color{black}

\begin{table}[t!]
\centering
\begin{tabularx}{0.455\textwidth}{|X|X|X|}\hline
Decay channel  & Current bound & Future sensitivity \\
& $\tau_p$ [yrs] &  $\tau_p$ [yrs]
\\\hline\hline
$p \rightarrow K^+\, \overline{\nu}$  &  $8.0 \times 10^{33}$ \cite{Okumura:2020xfs} &  $3.2\times 10^{34}$ \cite{Hyper-Kamiokande:2018ofw} \\
$p \rightarrow \pi^+\, \overline{\nu}$  &  $3.9 \times 10^{32}$ \cite{Super-Kamiokande:2013rwg} &  -  \\\hline\hline
$p \rightarrow \pi^0\, e^+$  &  $2.4 \times 10^{34}$ \cite{Super-Kamiokande:2020wjk} &  $7.8\times 10^{34}$ \cite{Hyper-Kamiokande:2018ofw}   \\
$p \rightarrow \pi^0\, \mu^+$  &  $1.6 \times 10^{34}$ \cite{Super-Kamiokande:2020wjk} &  $7.7\times 10^{34}$ \cite{Hyper-Kamiokande:2018ofw}    \\
$p \rightarrow K^0\, e^+$  &  $1.1 \times 10^{33}$ \cite{Brock:2012ogj} &  -  \\ 
$p \rightarrow K^0\, \mu^+$  &  $3.6 \times 10^{33}$ \cite{Super-Kamiokande:2022egr} &  -  \\
$p \rightarrow \eta^0\, e^+$  &  $1.0 \times 10^{34}$ \cite{Super-Kamiokande:2017gev} &  $4.3 \times 10^{34}$ \cite{Hyper-Kamiokande:2018ofw}  \\
$p\rightarrow \eta^0\,\mu^+$  &  $4.7 \times 10^{33}$ \cite{Super-Kamiokande:2017gev} &  $4.9 \times 10^{34}$ \cite{Hyper-Kamiokande:2018ofw}   \\\hline
\end{tabularx}
\caption{Current experimental bounds and future sensitivities (for 10 years of runtime) for different two-body proton decay channels. }\label{tab:nucleon_decay}
\vspace{3mm}
\begin{tabular}{|l|l|l|}\hline
Process  & Present bound & Future sensitivity  
\\\hline\hline
BR($\mu\to e\gamma$) & $4.2\times 10^{-13}$\ \cite{MEG:2016leq} & $6\times 10^{-14}$\ \cite{Baldini:2013ke}
\\
BR($\tau\to e\gamma$) & $3.3\times 10^{-8}$\ \cite{BaBar:2009hkt} & $\sim 10^{-9}$\ \cite{Aushev:2010bq}
\\
BR($\tau\to\mu\gamma)$ & $4.4\times 10^{-8}$\ \cite{BaBar:2009hkt} & $\sim 10^{-9}$\ \cite{Aushev:2010bq}
\\\hline\hline
BR($\mu\to e e e)$ & $1.0\times 10^{-12}$\ \cite{SINDRUM:1987nra} & $\sim 10^{-16}$\ \cite{Blondel:2013ia}
\\
BR($\tau\to e e e)$ & $2.7\times 10^{-8}$\ \cite{Hayasaka:2010np} & $\sim 10^{-9}$\ \cite{Aushev:2010bq}
\\
BR($\tau\to \mu\mu\mu)$ & $2.1\times 10^{-8}$\ \cite{Hayasaka:2010np} & $\sim 10^{-9}$\ \cite{Aushev:2010bq}
\\
BR($\tau^-\to e^-\mu\mu$) & $2.7\times 10^{-8}$\ \cite{Hayasaka:2010np} & $\sim 10^{-9}$\ \cite{Aushev:2010bq}
\\
BR($\tau^-\to \mu^-ee$) & $1.8\times 10^{-8}$\ \cite{Hayasaka:2010np} & $\sim 10^{-9}$\ \cite{Aushev:2010bq}
\\
BR($\tau^-\to e^+\mu^-\mu^-$) & $1.7\times 10^{-8}$\ \cite{Hayasaka:2010np} & $\sim 10^{-9}$\ \cite{Aushev:2010bq}
\\
BR($\tau^-\to \mu^+e^-e^-$) & $1.5\times 10^{-8}$\ \cite{Hayasaka:2010np} & $\sim 10^{-9}$\ \cite{Aushev:2010bq}
\\
\hline\hline
CR($\mu \textrm{Au}\to e \textrm{Au})$ & $7\times 10^{-13}$\ \cite{SINDRUMII:2006dvw} & $-$
\\
CR($\mu \textrm{Ti}\to e \textrm{Ti})$ & $4.3\times 10^{-12}$\ \cite{SINDRUMII:1993gxf} & $\sim 10^{-18}$\ \cite{unPUB}
\\
CR($\mu \textrm{Al}\to e \textrm{Al})$ & $-$ & $10^{-15}-10^{-18}$\ \cite{Pezzullo:2017iqq}
\\
\hline
\end{tabular}
\caption{Current experimental constraints and future sensitivities for various lepton violating processes; all at the 90\% confidence level.}\label{tab:cLFV}
\end{table}

\begin{figure*}[ht!]
     \raisebox{-1mm}{\includegraphics[width=0.47\textwidth]{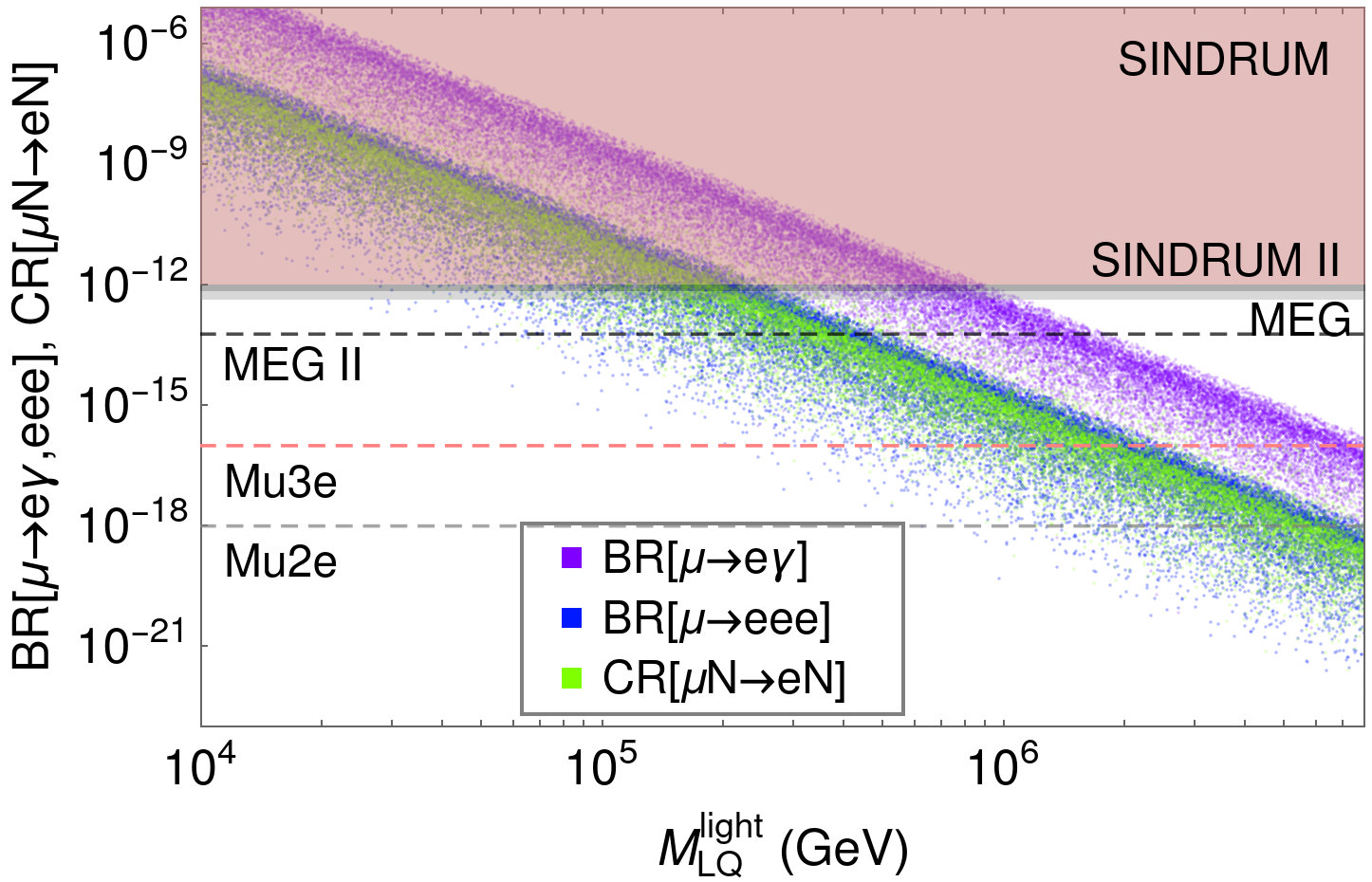}}
    \hfill
    \includegraphics[width=0.47\textwidth]{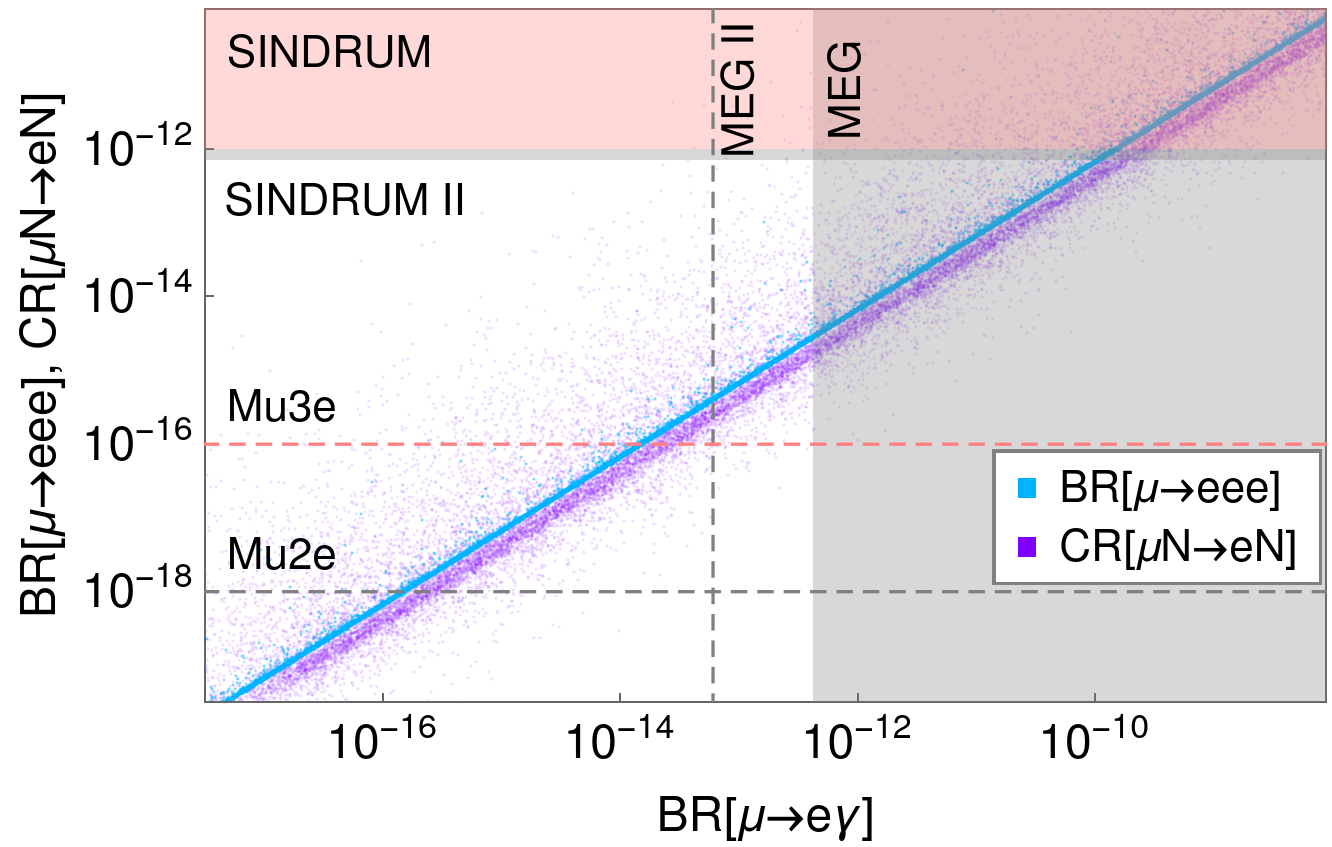}
    \caption{Left panel: Correlation between various cLFV processes and the LQ mass. Right panel: Correlation between various cLFV processes amongst each other. Current experimental constraints are represented by filled boxes, while future sensitivities are indicated by dashed lines. See text for details.}\label{fig:LFV}
\end{figure*}

\subsection{Lepton flavor violation}
The Yukawa couplings Eqs.~\eqref{eq:nu-yuk-1}-\eqref{eq:nu-yuk-4} that appear in the neutrino mass formula Eq.~\eqref{eq:nu-mass} naturally lead to lepton flavor violating processes. Among the three different types of leptoquarks, the couplings of $\chi^{\pm 1/3}$ and $\chi^{\pm 5/3}$ are severely constrained owing to chirality-enhanced lepton flavor violating processes as a result of the simultaneous presence of both left- and right-chiral interactions with fermions. On the other hand, cLFV processes lead by $\chi^{\pm 2/3}$ leptoquark are comparatively  suppressed. In our analysis, we take into account all the important two-body and three-body decays $\ell_\alpha\to\ell_\beta \gamma$ and $\ell^-_\alpha\to \ell^-_\beta\ell^-_\gamma\ell^+_\gamma$, as well as $\mu\to e$ conversion in nuclei. For the computation of these processes, we follow Ref.~\cite{Julio:2022ton} and the references therein.  Moreover, Table~\ref{tab:cLFV} summarizes all current experimental bounds and sensitives of upcoming experiments. As we will shown below, the present lepton flavor violating constraints typically put a lower bound of $M_\mathrm{LQ}\gtrsim 10^5$ GeV on the mass of the leptoquarks.

\subsection{Numerical analysis and results}
For our numerical analysis we work in the basis in which the charged lepton mass matrix is diagonal. We then parameterize the Yukawa matrices $Y_A,Y_C,(Y_B+Y_B^T)$, and $(Y_D-Y_D^T)$ in terms of the charged fermion masses and the unitary rotation matrices $U_L,U_R,D_L$, and $D_R$: 
\begin{align}
    &Y_A=\frac{1}{2v_5}\left(M_E^{\mathrm{diag}}+3D_LM_D^{\mathrm{diag}}D_R^\dagger\right),
    \\
    &Y_C=\frac{\sqrt{3}}{\sqrt{2}v_{45}}\left(M_E^{\mathrm{diag}}-D_LM_D^{\mathrm{diag}}D_R^\dagger\right),
    \\
    &Y_B+Y_B^T=\frac{1}{2\sqrt{2}v_5}\left(U_LM_U^{\mathrm{diag}}U_R^\dagger+U_R^\ast M_U^{\mathrm{diag}} U_L^T\right),
    \\
    &Y_D-Y_D^T=\frac{\sqrt{3}}{2v_{45}}\left(U_LM_U^{\mathrm{diag}}U_R^\dagger-U_R^\ast M_U^{\mathrm{diag}} U_L^T\right).
\end{align}
Note that we parametrize the up-type quark left-mixing rotation matrix as $U_L=D_L \textrm{diag}(e^{i\beta_1^u},e^{i\beta_2^u},1)V_{\mathrm{CKM}}^T$, where $V_{\mathrm{CKM}}$ is the Cabibbo-Kobayashi-Maskawa (CKM) matrix.
This parametrization is derived by solving Eqs.~\eqref{eq:MD}-\eqref{eq:MU} for the Yukawa matrices. It has the advantage that the charged fermion masses and CKM mixing angles can be used as input parameters (which we take from Ref.~\cite{Babu:2016bmy}), hence these quantities do not require to be fitted. Therefore, we scan over the rest of the free parameters in the rotation matrices. The neutrino mass matrix is in this basis given as
\onecolumngrid
\begin{align}
    M_N=\frac{\left(\frac{3}{2}\right)^{\frac{3}{2}}g^2}{(4\pi)^4}&\bigg\{\left(M_E^\mathrm{diag}+3M_D^\dagger\right)M_U^\ast\left(M_E^\mathrm{diag}-M_D\right)\frac{\hat{I}}{v_5v_{45}}
    \nonumber\\
    &-M_E^\mathrm{diag}\left(M_U^\ast+M_U^\dagger\right)\left(M_E^\mathrm{diag}-M_D\right)\frac{\tilde{I}}{v_{45}^2}
    +M_E^\mathrm{diag}\left(M_E^\mathrm{diag}+3M_D\right)\left(M_U^\ast-M_U^\dagger\right)\frac{\bar I}{2v_5v_{45}}\bigg\}.
\end{align}
\twocolumngrid

We use these free parameters discussed above together with the ones defined in Eq.~\eqref{eq4} to fit the neutrino masses and PMNS mixings (we take the experimental values from Refs.~\cite{Esteban:2020cvm, NUFIT}), while making sure that all proton decay and cLFV constraints are satisfied, and while choosing the intermediate-scale particle masses such that the gauge couplings do unify. Using a differential evolution algorithm we compute 12 different benchmark points; all having a negligible total $\chi^2$.  Starting from each benchmark point and applying an adaptive Metropolis-Hastings algorithm we perform a Markov-chain-Monte-Carlo (MCMC) analysis. For each chain we compute 1.5 million points giving us 18 million points in total. 

It must be noted that we only find good fit points for the case of normal neutrino mass ordering. This can be understood from the fact that the neutrino mass matrix is highly correlated to the charged fermion mass matrices. Consequentially, the neutrino mass matrix mirrors the normal mass hierarchy of the mass matrices of the charged fermions.

We find the most constraining cLFV processes to be $\mu\rightarrow e\gamma$, $\mu\rightarrow eee$, and $\mu-e$ conversion in nuclei. We, therefore, present in Fig.~\ref{fig:LFV} the correlation between these three processes and the leptoquark mass scale $M_{\mathrm{LQ}}^\mathrm{light}$ (left panel) as well as the correlation of these three processes amongst each other (right panel). The current bound on the process $\mu\rightarrow e\gamma$ typically constrains $M_{\mathrm{LQ}}^\mathrm{light}\gtrsim 10^5$~GeV. Future sensitivities on the processes $\mu\rightarrow eee$ and $\mu-e$ conversion in nuclei have the potential to increase this bound by an order of magnitude if experiments happen not to observe any cLFV process. Moreover, we find a high correlation between the cLFV processes $\mu\rightarrow e\gamma$ and $\mu\rightarrow eee$ that can be looked for in future searches. We find BR($\mu\rightarrow e\gamma$)/BR($\mu\rightarrow eee)$ to lie within $(108,173)$ at the 2$\sigma$ confidence level (see Fig.~\ref{fig:LFV}).

\begin{figure*}[t!]
\raisebox{0.5mm}{
    \includegraphics[width=0.465\textwidth]{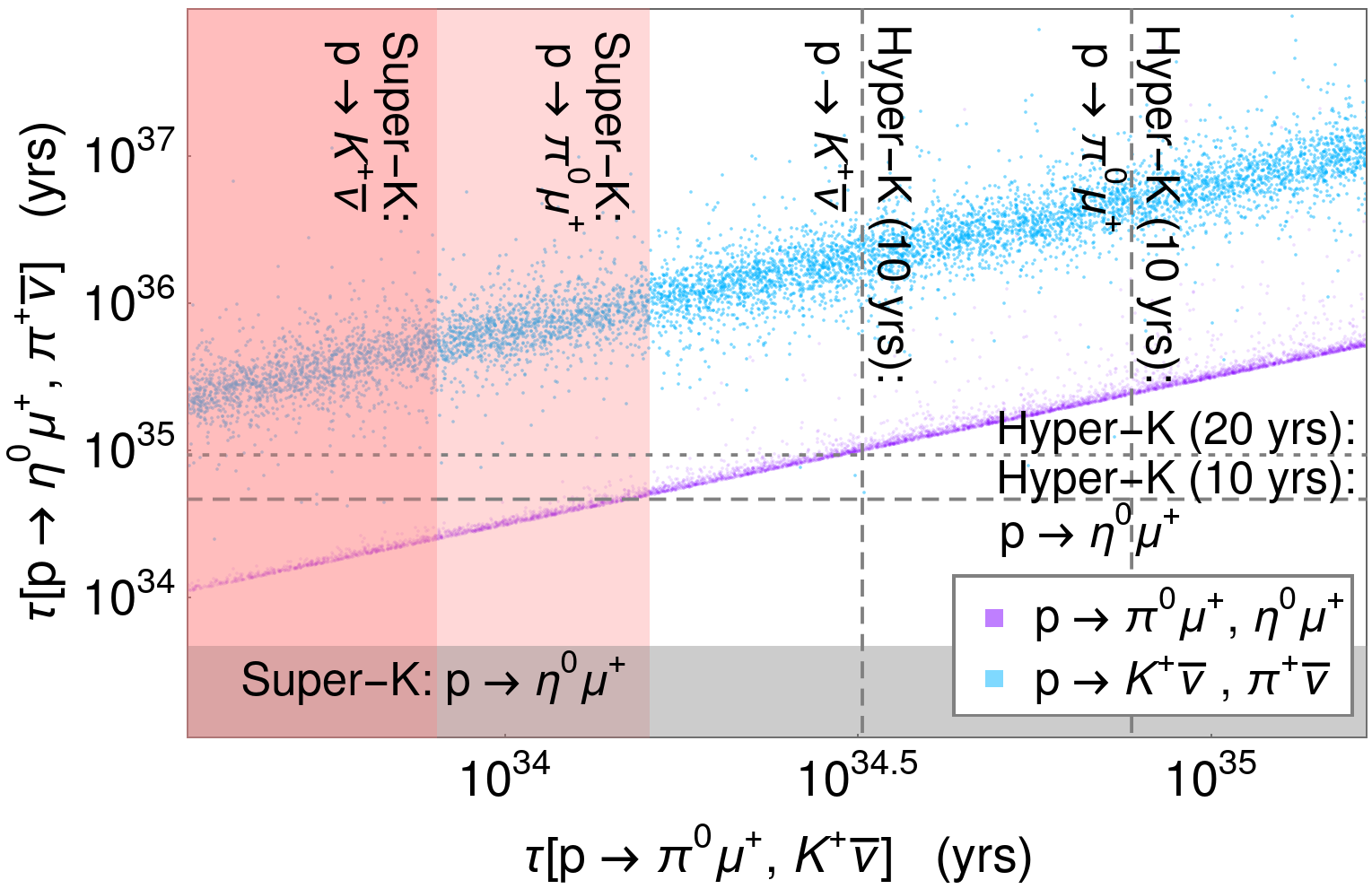}
    }
    \hfill
    \includegraphics[width=0.48\textwidth]{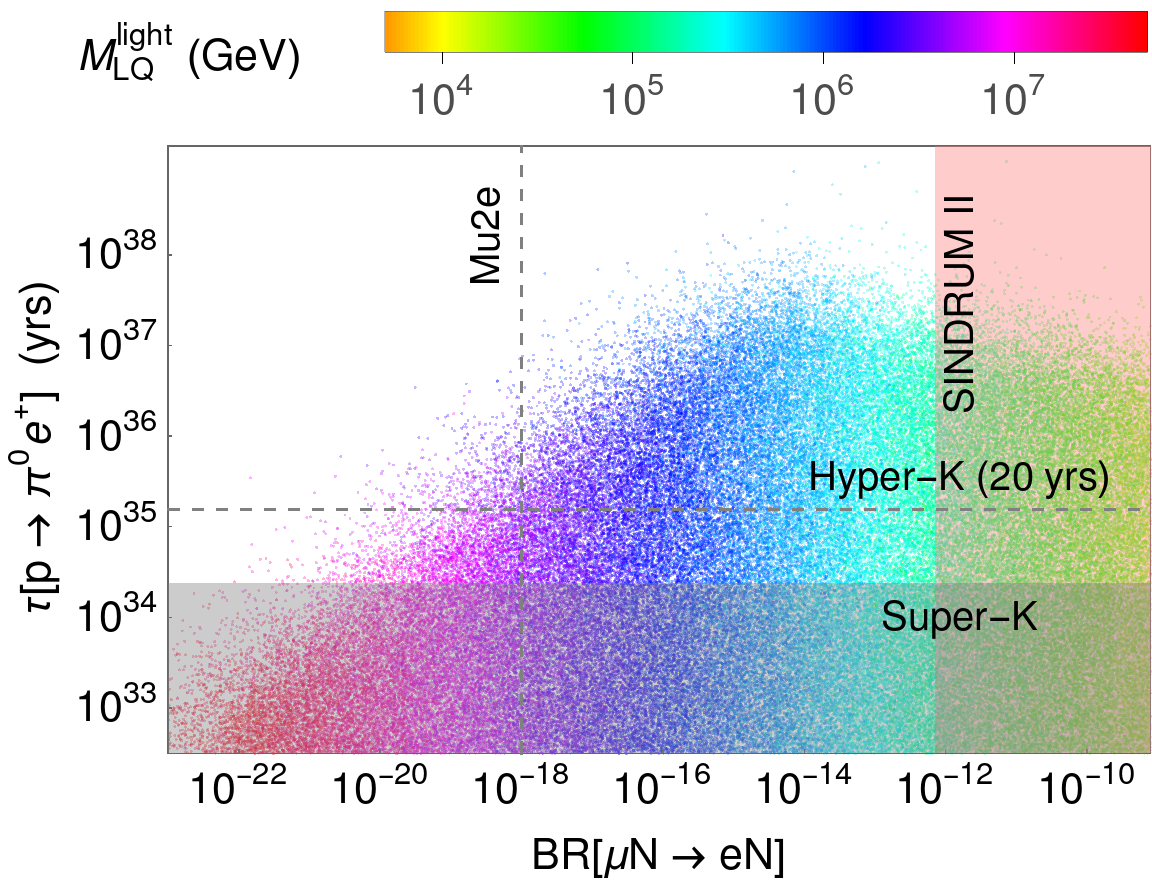}
    \caption{Left panel: Correlation between various proton decay channels. The violet points show the correlation between the partial proton lifetimes $\tau(p\rightarrow \pi^0\mu^+)$ and $\tau(p\rightarrow \eta^0\mu^+)$, while the cyan points indicate the correlation between $\tau(p\rightarrow K^+\overline{\nu})$ and $\tau(p\rightarrow \pi^+\overline{\nu})$.   
    Right panel: Correlation between cLFV and proton decay.  See text for details. }
    \label{fig:lfv_vs_pd}
\end{figure*}

For the gauge mediated proton decay, we find the channels $p\rightarrow \pi^0\ell^+$ to be typically dominant. But also proton decay in the channels $p\rightarrow \eta\ell^+$ or $p\rightarrow K^+\overline{\nu}$ may be observed. Interestingly, we find a high correlation between the decay channels $p\rightarrow \pi^0\mu^+$ and $p\rightarrow \eta^0\mu^+$ that can be looked for at Hyper-Kamiokande. At the $2\sigma$ confidence level, we find the ratio of partial lifetimes $\tau(p\rightarrow \pi^0\mu^+)/\tau(p\rightarrow \eta^0\mu^+)$ to lie within $(0.12,0.33)$. This correlation is depicted in the left panel of Fig.~\ref{fig:lfv_vs_pd}. 

In the case of proton decay mediated by the scalar leptoquark $\Sigma_2$,  we find amongst the usual channels that the decays $p\rightarrow K^0\ell^+$ are enhanced and could be tested if $\Sigma_2$ resides around $10^{12}$~GeV. Considering the leptoquark $\Sigma_5$, the dominant decay channel is $p\rightarrow K^+\overline{\nu}$. It can be looked for if $\Sigma_5$ resides around $10^{12}$~GeV. The same decay channel is also the dominant one mediated by the $\chi^{\pm 1/3}_a$ states. Moreover, for $\Sigma_5$-mediated proton decay, we find a somewhat interesting correlation between the two decay channels with antineutrinos in the final state. We findings dictate that the ratio of partial lifetimes $\tau(p\rightarrow K^+\overline{\nu})/\tau(p\rightarrow \pi^+\overline{\nu})$ to be within $(0.0052, 0.0467)$ at the $2\sigma$ confidence level. This correlation is visualized in the left panel of Fig.~\ref{fig:lfv_vs_pd}. 

Moreover, in the minimal scenario, where we freely vary the masses $M_{\Phi_2},M_{\Phi_3},M_{\Sigma_7},M_{\eta_5},M_{\mathrm{LQ}}^\mathrm{light}$ and $M_{\mathrm{LQ}}^\mathrm{heavy}$ for a gauge coupling unification analysis, while assuming that all other states reside around the GUT scale, there is additional correlation between lepton flavor violating processes and gauge mediated proton decay. This is due to the fact that the GUT scale depends on the choice of the leptoquark mass scale, $M_{\mathrm{LQ}}^\mathrm{light}$. We illustrate this correlation by considering the cLFV process muon to electron conversion in nuclei and the proton decay channel $p\rightarrow \pi^0e^+$ in the right panel of Fig.~\ref{fig:lfv_vs_pd}.
Interestingly, if the leptoquark mass scale $M_{\mathrm{LQ}}^\mathrm{light}$ gets larger than $10^7$~GeV gauge mediated proton decay violates the present Super-Kamiokande bound. On the other hand, a leptoquark mass scale $M_{\mathrm{LQ}}^\mathrm{light}$ below $10^5$~GeV is ruled out by current cLFV constraints.
If not observed, future sensitivities on cLFV processes have the potential to increase this bound on the leptoquark mass scale by an order of magnitude.

From our comprehensive analysis, we conclude that the proposed model has exciting feature through which it can be simultaneously tested by a synergy of low energy searches looking for proton decay and cLFV.

Before concluding this section, we emphasize again that, in contrast to other SU(5) GUT-based neutrino mass models, our proposal stands out for its use of colored states, specifically scalar leptoquarks, to generate neutrino mass. Unlike existing models that rely on non-colored multiplets (all  tree-level neutrino mass models Refs.~\cite{Dorsner:2005fq,Dorsner:2005ii,Bajc:2006ia,Dorsner:2006hw,Dorsner:2007fy,Antusch:2021yqe,Antusch:2022afk,Calibbi:2022wko,Antusch:2023kli,Antusch:2023mqe} as well as radiative neutrino mass models Refs.~\cite{Wolfenstein:1980sf,Barbieri:1981yw,Perez:2016qbo,Kumericki:2017sfc,Saad:2019vjo,Dorsner:2019vgf,Dorsner:2021qwg,Antusch:2023jok,Dorsner:2024jiy}), ours operates at two-loop order, resulting in a unique implication: a set of leptoquarks remains light. Furthermore, the couplings of these leptoquarks with the SM fermions are determined by the requirements of reproducing observed fermion mass spectrum. Consequently, our model diverges from existing ones in its phenomenological implications, particularly in inducing enhanced lepton flavor violation by colored mediators. Observations of rare lepton flavor-violating processes, exploration of correlations among different modes, and further associations between LFV and specific proton decay channels could offer distinctive signatures for our model. In particular, for gauge mediated proton decay, our model predicts a high correlation between the proton decay channels $p\rightarrow\pi^0\mu^+$ and $p\rightarrow\eta^0\mu^+$. The ratio of their partial lifetimes lies at the 2$\sigma$ confidence level within $(0.12,0.33)$. Another unique signature of our model is the enhancement of the proton decay channels $p\rightarrow K^0\ell^+$ via scalar leptoquarks.

Furthermore, within our scenario, color sextets $(\overline{6}, 1, 2/3)$ and $(\overline{6}, 1, -1/3)$, which are part of $\eta_5$ originating from $40_H$, and color octets $(8,1,0)$ and $(8,1,1)$, which are part of $\Sigma_7$, could reside at low energies. If their masses are close to the TeV scale, they can be efficiently pair-produced through gluon-fusion at the LHC. Once produced, these sextets, either having zero or very suppressed Yukawa couplings, would behave like long-lived colored states, namely R-hadrons~\cite{Farrar:1978xj}. These practically stable colored states would decay outside the detectors and behave like long-lived gluinos or squarks. The LHC has extensively searched for these long-lived super-partners, and the current bounds on these states translate into $m \gtrsim 1250$ GeV~\cite{ATLAS:2019gqq} and $m \gtrsim 1800$ GeV~\cite{CMS:2020iwv} for $Q_\mathrm{em} = \pm 1/3$ and $Q_\mathrm{em} = \pm 2/3$, respectively. Although the color octets are produced in a similar way, due to their sizable Yukawa couplings, they dominantly decay to third-generation quarks. The charged partner provides a stronger bound since, unlike its neutral partner, it cannot decay to gluons. Consequently, searches for these states at the LHC focus on the signal $pp \to tb\overline{t}\overline{b}$. Analysis of the current LHC data provides a lower mass bound of $\gtrsim 800$ GeV for these particles~\cite{Miralles:2019uzg}.  \color{black}

\section{Conclusions}\label{conclusion}
Scalar leptoquarks emerge as natural constituents within unified theories, providing a compelling solution to one of the Standard Model's critical puzzles--the presence of non-zero neutrino masses. In this study, we proposed a SU(5) grand unified model, in which neutrino masses are generated at the two-loop level through scalar leptoquark propagation. This unified approach intricately links the masses and mixing of charged fermions and neutrinos, governed by a unified set of Yukawa interactions. These exotic particles not only illuminate the process behind neutrino mass generation but also contribute to the unification of gauge couplings and may result in significant rates of lepton flavor violation, opening distinct pathways for experimental verification. The masses of a set of leptoquarks are predicted to lie several orders of magnitude below the unification scale from the requirement of reproducing the correct neutrino mass scale. Furthermore, to maximize the unification scale, new physics states such as color sextet and octet scalars should be at the TeV scale, placing them within reach of colliders. The diverse roles played by the leptoquarks highlight the elegance and predictive ability of the proposed unified model.

\section*{Acknowledgments}
We thank the anonymous referee for prompting us to investigate loop-mediated proton decays.   S.S.\ would like to thank I. Dor\v{s}ner and J. Julio for discussion. 

\onecolumngrid
\appendix
\section{Gauge mediated proton decay}\label{sec:Apx-01}
The complete formula for the gauge mediated two-body proton decay widths are \cite{FileviezPerez:2004hn,Nath:2006ut}
\begingroup
\allowdisplaybreaks
\begin{align}
 \label{eq:Gamma p->K nu,gauge}
&\Gamma (p \to K^+ \bar{\nu})= 
\frac{(m_p-m_{K^+})^2}{8\pi m_p^3}A_L^2 
 \sum_i  \left|  
A_{SR}  \matrixel{K^+}{(us)_R d_L}{p}  c(\nu_i,d,s^c)  
+ A_{SL}  \matrixel{K^+}{(ud)_R s_L}{p}  c(\nu_i,s,d^c)  \right|^2 \;,
    \\ \label{eq:Gamma p->pi nu,gauge}
&\Gamma (p \to \pi^+ \bar{\nu})= 
\frac{(m_p-m_{\pi^+})^2}{8\pi m_p^3}A_L^2A_{SR}^2 |\matrixel{\pi^+}{(ud)_R d_L}{p}|^2 \sum_i |c(\nu_i,d,d^c)|^2\;,
\\ \label{eq:Gamma p->pi e,gauge}
&\Gamma (p \to \pi^0 e^+_\alpha)= 
\frac{(m_p-m_{\pi^0})^2}{8\pi m_p^3}A_L^2
\left(A_{SL}^2|\langle\pi^0|(ud)_Lu_L|p\rangle c(e_\alpha^c,d)|^2+A_{SR}^2|\langle\pi^0|(ud)_Ru_L|p\rangle c(e_\alpha,d^c)|^2\right),
\\ \label{eq:Gamma p->K e,gauge}
    &\Gamma (p \to K^0 e^+_\alpha)= 
\frac{(m_p-m_{K^0})^2}{8\pi m_p^3}A_L^2
\left(A_{SR}^2|\langle K^0|(us)_Ru_L|p\rangle c(e_\alpha,s^c)|^2+A_{SL}^2|\langle K^0|(ud)_Lu_L|p\rangle c(e_\alpha^c,s)|^2\right),
    \\ \label{eq:Gamma p->eta e,gauge}
    &\Gamma (p \to \eta e^+_\alpha)= 
\frac{(m_p-m_{\eta})^2}{8\pi m_p^3}A_L^2
\left(A_{SL}^2|\langle\eta|(ud)_Lu_L|p\rangle c(e_\alpha^c,d)|^2+A_{SR}^2|\langle\eta|(ud)_Ru_L|p\rangle c(e_\alpha,d^c)|^2\right).
\end{align}
\endgroup
Here, $A_L=1.2$ denotes the long-range coefficient \cite{Nihei:1994tx}, while $A_{SL(R)}$ are the short-distance coefficients defined as 
\begin{align}
    &A_{SL(R)}=\prod_{i=1,2,3}\prod_{I}^{M_Z\leq M_I\leq M_\textrm{GUT}}\left(\frac{\alpha_i(M_{I+1})}{\alpha_i(M_I)}\right)^{\frac{\gamma_{L(R)i}}{b_i^\textrm{SM}+\sum_J^{M_Z\leq M_J\leq M_\textrm{GUT}}b_i^J}},
    \;
    \text{with $\gamma_{L(R)i}=\left(23(11)/20,9/4,2\right)$.}
\end{align}
Moreover, $m_p=938.33$\ MeV, $m_\pi=134$\ MeV, $m_\eta=548$\ MeV, and $m_K=493.677$\ MeV are the proton, pion, eta meson, and kaon mass, respectively. 
The coefficients are defined as \cite{Buras:1977yy,Ellis:1979hy,Wilczek:1979hc}
\begin{align}
    &c(e_\alpha^c,d_\beta)=\frac{g_{\textrm{GUT}}^2}{2M_{\textrm{GUT}}^2}\left((U_R^\dagger U_L^\ast)_{11}(E_R^\dagger D_L^\ast)_{\alpha\beta}+(E_R^\dagger U_L^\ast)_{\alpha 1}(U_R^\dagger D_L^\ast)_{1\beta}\right)\ ,\\
    &c(e_\alpha,d^c_\beta)=\frac{g_{\textrm{GUT}}^2}{2M_{\textrm{GUT}}^2}(U_R^\dagger U_L^\ast)_{11}(E_L^\dagger D_R^\ast )_{\alpha\beta}\ ,\\
    &c(\nu_l,d_\alpha,d^c_\beta)=\frac{g_{\textrm{GUT}}^2}{2M_{\textrm{GUT}}^2}(U_R^\dagger D_L^\ast)_{1\alpha}(D_R^\dagger N)_{\beta l}\ .
\end{align} 
Finally, we take the matrix elements from Ref.~\cite{Aoki:2017puj}, e.g.~$\matrixel{\pi^0}{(ud)_Lu_L}{p}=+0.134(5)(16) \text{ GeV}^2$.

\section{Scalar mediated proton decay}\label{sec:Apx-02}

The decay widths for scalar mediated two-body proton decay read  \cite{FileviezPerez:2004hn,Nath:2006ut}
\begingroup
\allowdisplaybreaks
\begin{align}
\nonumber
&\Gamma (p \to K^+ \bar{\nu})= 
\frac{(m_p-m_{K^+})^2}{32\pi m_p^3}A_L^2
   \sum_i  \left|  
A_{SL}\left(\langle K^+|(us)_Ld_L|p\rangle a(s,d,\nu_i)+\langle K^+|(ud)_Ls_L|p\rangle a(d,s,\nu_i)\right) \right.\  
\\
 \label{eq:Gamma p->K nu,scalar}
 &\hspace{5cm}  \left.\ +\, A_{SR}\left(\langle K^+|(us)_Rd_L|p\rangle a(d,s^c,\nu_i)+\langle K^+|(ud)_Rs_L|p\rangle a(s,d^c,\nu_i)\right) \right|^2  \;,
    \\ \label{eq:Gamma p->pi nu,scalar}
&\Gamma (p \to \pi^+ \bar{\nu})= 
\frac{(m_p-m_{\pi^+})^2}{32\pi m_p^3}A_L^2 
\sum_i \left|A_{SL}\matrixel{\pi^+}{(ud)_L d_L}{p}a(d,d,\nu_i)+A_{SR}\matrixel{\pi^+}{(ud)_R d_L}{p}a(d,d^c,\nu_i)\right|^2\;,
\\ \label{eq:Gamma p->pi e,scalar}
\nonumber
&\Gamma (p \to \pi^0 e^+_\alpha)= 
\frac{(m_p-m_{\pi^0})^2}{32\pi m_p^3}A_L^2
\left( 
\left|A_{SL}\langle\pi^0|(ud)_Lu_L|p\rangle a(e_\alpha,d)+A_{SR}\langle\pi^0|(ud)_Ru_L|p\rangle a(e_\alpha,d^c)\rangle\right|^2\right.\
\\
&\hspace{5cm}
\left.\ +\left|A_{SR}\langle\pi^0|(ud)_Lu_L|p\rangle a(e_\alpha^c,d^c)+A_{SL}\langle\pi^0|(ud)_Ru_L|p\rangle a(e_\alpha^c,d)\right|^2\right)
,
\\ 
\nonumber
    &\Gamma (p \to K^0 e^+_\alpha)= 
\frac{(m_p-m_{K^0})^2}{64\pi m_p^3}A_L^2
\left(\left|A_{SR}\langle K^0|(us)_Ru_L|p\rangle a(e_\alpha,s^c)+A_{SL}\langle K^0|(us)_Lu_L|p\rangle a(e_\alpha,s)\right.\ \right.\
\\\nonumber
&\hspace{5cm}\left.\ -\,A_{SL}\langle K^0|(us)_Ru_L|p\rangle a(e_\alpha^c,s)-A_{SR}\langle K^0|(us)_Lu_L|p\rangle a(e_\alpha^c,s^c)\right|^2
\\\nonumber
&\hspace{5.15cm}+\left|A_{SR}\langle K^0|(us)_Ru_L|p\rangle a(e_\alpha,s^c)+A_{SL}\langle K^0|(us)_Lu_L|p\rangle a(e_\alpha,s)\right.\
\\\label{eq:Gamma p->K e,scalar}
 &\hspace{4.85cm}\left.\ \left.\ +\,A_{SL}\langle K^0|(us)_Ru_L|p\rangle a(e_\alpha^c,s)+A_{SR}\langle K^0|(us)_Lu_L|p\rangle a(e_\alpha^c,s^c) \right|^2\right)
,
    \\ \label{eq:Gamma p->eta e,scalar}
\nonumber
&\Gamma (p \to \eta e^+_\alpha)= 
\frac{(m_p-m_{\eta})^2}{32\pi m_p^3}A_L^2
\left( 
\left|A_{SL}\langle\eta|(ud)_Lu_L|p\rangle a(e_\alpha,d)+A_{SR}\langle\eta|(ud)_Ru_L|p\rangle a(e_\alpha,d^c)\rangle\right|^2\right.\
\\
&\hspace{5cm}
\left.\ +\left|A_{SR}\langle\eta|(ud)_Lu_L|p\rangle a(e_\alpha^c,d^c)+A_{SL}\langle\eta|(ud)_Ru_L|p\rangle a(e_\alpha^c,d)\right|^2\right).
\end{align}
\endgroup
\underline{Tree level:}\\
For $\Sigma_2^{-1/3}$ the coefficients are in our model \cite{Dorsner:2012nq}
\begin{align}
    &a(e_\alpha,d_\beta)=a(e_\alpha^c,d_\beta)=a(d_\alpha,d_\beta,\nu_i)=0,\\
    &a(e_\alpha,d_\beta^c)=\frac{1}{M_{\Sigma_2}^2}\left(D_R^\dagger \frac{Y_C^\dagger}{2}U_R^\ast\right)_{\beta 1}\left(U_L^\dagger \frac{-Y_C}{2}e_L^\ast\right)_{1\alpha},\\
    &a(e_\alpha^c,d_\beta^c)=\frac{1}{M_{\Sigma_2}^2}\left(D_R^\dagger \frac{Y_C^\dagger}{2}U_R^\ast\right)_{\beta 1}\left(E_R^\dagger\sqrt{2}(Y_D^\ast-Y_D^\dagger)U_R^\ast\right)_{\alpha 1},\\
    &a(d_\alpha,d_\beta^c,\nu_i)=\frac{1}{M_{\Sigma_2}^2}\left(D_R^\dagger \frac{Y_C^\dagger}{2}U_R^\ast\right)_{\beta 1}\left(D_L^\dagger \frac{Y_C}{2}N^\ast\right)_{\alpha i}.
\end{align}
For $\Sigma_5^{-1/3}$ we find 
\begin{align}
    &a(e_\alpha^c,d_\beta)=a(e_\alpha,d_\beta^c)=a(e_\alpha^c,d_\beta^c)=a(d_\alpha,d_\beta^c,\nu_i)=0,\\
    &a(e_\alpha,d_\beta)=\frac{1}{M_{\Sigma_5}^2}\left(U_L^\dagger(Y_D^T-Y_D)D_L^\ast\right)_{1\beta}\left(U_L^\dagger\frac{Y_C}{\sqrt{2}}E_L^\ast\right)_{1\alpha},\\
    &a(d_\alpha,d_\beta,\nu_i)=\frac{1}{M_{\Sigma_5}^2}\left(U_L^\dagger(Y_D^T-Y_D)D_L^\ast\right)_{1\alpha}\left(D_L^\dagger \frac{Y_C}{\sqrt{2}}N^\ast\right)_{\beta_i}.
\end{align}
For $\Sigma_5^{+2/3}$ the coefficients read
\begin{align}
    &a(e_\alpha,d_\beta)=a(e_\alpha^c,d_\beta)=a(e_\alpha,d_\beta^c)=a(e_\alpha^c,d_\beta^c)=a(d_\alpha,d_\beta^c,\nu_i)=0,\\
    &a(d_\alpha,d_\beta,\nu_i)=\frac{2}{M_{\Sigma_5}^2}\left(U_L^\dagger(-Y_C)N^\ast\right)_{1i}\left(D_L^\dagger\sqrt{2}(Y_D-Y_D^T)D_L^\ast\right)_{\beta\alpha}.
\end{align}
Finally, the coefficients for $\phi_2^{-1/3}$ are given by (from which we can easily get the relevant coefficients for $\chi^{\pm 1/3}_a$ by multiplying with $\sin\theta$ ($\cos\theta$) for $a=1$ ($a=2$))
\begingroup
\allowdisplaybreaks
\begin{align}
    &a(e_\alpha,d_\beta)=\frac{1}{M_{\phi_2}^2}\left(-U_L^\dagger 2(Y_B+Y_B^T)D_L^\ast\right)_{1\beta}\left(U_L^\dagger \frac{-Y_A}{\sqrt{2}}E_L^\ast\right)_{1\alpha},\\
    &a(e_\alpha^c,d_\beta)=\frac{1}{M_{\phi_2}^2}\left(-U_L^\dagger 2(Y_B+Y_B^T)D_L^\ast\right)_{1\beta}\left(E_R^\dagger 2(Y_B^\ast+Y_B^\dagger)U_R^\ast\right)_{\alpha 1},\\
    &a(e_\alpha,d_\beta^c)=\frac{1}{M_{\phi_2}^2}\left(D_R^\dagger\frac{Y_A^\dagger}{\sqrt{2}}U_R^\ast\right)_{\beta 1}\left(U_L^\dagger \frac{-Y_A}{\sqrt{2}}E_L^\ast\right)_{1\alpha},\\
    &a(e_\alpha^c,d_\beta^c)=\frac{1}{M_{\phi_2}^2}\left(D_R^\dagger\frac{Y_A^\dagger}{\sqrt{2}}U_R^\ast\right)_{\beta 1}\left(E_R^\dagger 2(Y_B^\ast+Y_B^\dagger)U_R^\ast\right)_{\alpha 1},\\
    &a(d_\alpha,d_\beta,\nu_i)=\frac{1}{M_{\phi_2}^2}\left(-U_L^\dagger 2(Y_B+Y_B^T)D_L^\ast\right)_{1\alpha}\left(D_L^\dagger \frac{Y_A}{\sqrt{2}}N^\ast\right)_{\beta i},\\
    &a(d_\alpha,d_\beta^c,\nu_i)=\frac{1}{M_{\phi_2}^2}\left(D_R^\dagger \frac{Y_A^\dagger}{\sqrt{2}}U_R^\ast\right)_{\beta 1}\left(D_L^\dagger \frac{Y_A}{\sqrt{2}}N^\ast\right)_{\alpha i}.
\end{align}
\endgroup

\begin{figure}
    \centering
    \includegraphics[width=0.33\textwidth]{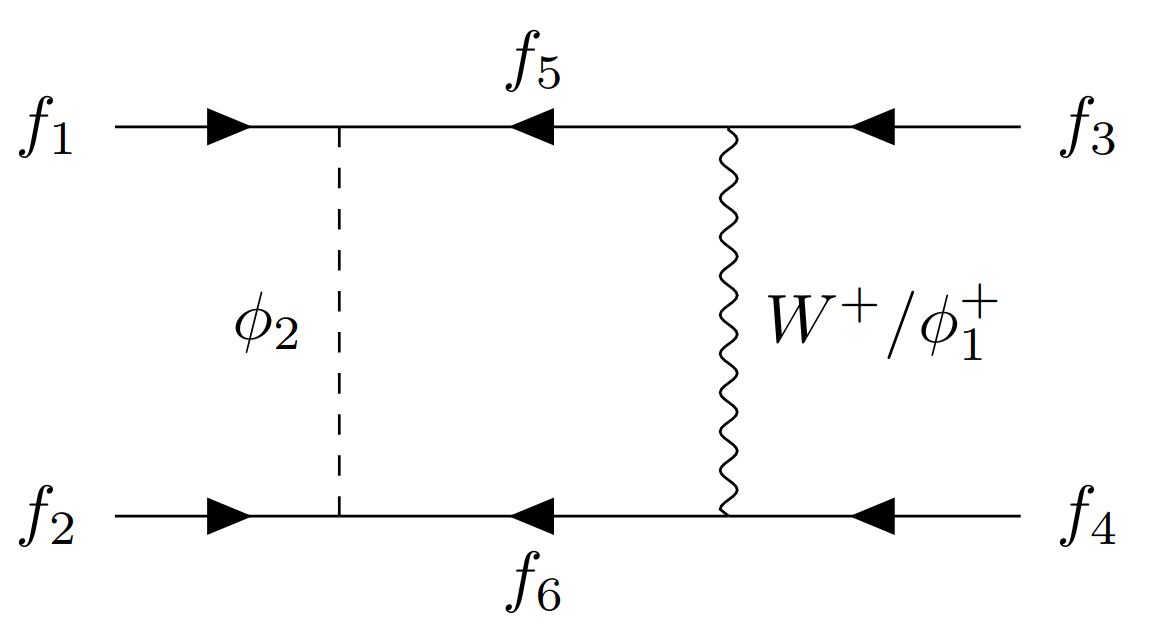}
    \caption{Representative box diagram for dimension six proton decay with a $W$ boson and a $\phi_2$ LQ running inside the loop.}
    \label{fig:pd_1loop}
\end{figure}

\noindent\underline{Loop level:}\\
The loop level contributions with a single $W$ boson running in the box diagram (see Fig.~\ref{fig:pd_1loop} for a representative diagram) to the coefficients for $\phi_2^{-1/3}$ are computed to be (the relevant coefficients for $\chi^{\pm 1/3}_a$ are obtained by multiplying with the corresponding mixing angles)
\begingroup
\allowdisplaybreaks
\begin{align}
    &a(e_\alpha,d^c_\beta)=a(e^c_\alpha,d^c_\beta)=0
    ,\\
    &a(e_\alpha,d_\beta)=
    4g_2^2I_2(m_{d_k},m_{\nu_l})
    \left(V^\ast_{\mathrm{CKM}}\right)_{1k} 
    \left(U_{\mathrm{PMNS}}\right)_{\alpha l}
    \left(D_L^\dagger\frac{Y_A}{\sqrt{2}}N^\ast\right)_{\beta l}
    \left(-U_L^\dagger 2(Y_B+Y_B^T)D_L^\ast\right)_{1k}
    \nonumber\\
    &\hspace{14mm} +4g_2^2 I_2(m_{d_k},m_{u_l})
    \left(V^\ast_{\mathrm{CKM}}\right)_{1k} 
    \left(V_{\mathrm{CKM}}\right)_{l\beta}
    \left(-U_L^\dagger 2(Y_B+Y_B^T)D_L^\ast\right)_{1k}
    \left(U_L^\dagger \frac{-Y_A}{\sqrt{2}}E_L^\ast\right)_{l\alpha}
    ,\\
    &a(e^c_\alpha,d_\beta)=g_2^2
    I_1(m_{d_k},m_{u_l}) 
    \left(V^\ast_{\mathrm{CKM}}\right)_{1k} 
    \left(V_{\mathrm{CKM}}\right)_{l\beta}
    \left(U_R^\dagger \frac{Y_A^\ast}{\sqrt{2}}D_R^\ast\right)_{1k}
    \left(U_R^\dagger 2(Y_B^\dagger+Y_B^\ast)E_R^\ast\right)_{l\alpha}
    ,\\
    &a(d_\alpha,d_\beta,\nu_i)=-\;4 g_2^2
    I_2(m_{d_k},m_{j_l})
     \left(V^\ast_{\mathrm{CKM}}\right)_{1k} 
    \left(V_{\mathrm{CKM}}\right)_{l\alpha}
    \left(-U_L^\dagger 2(Y_B+Y_B^T)D_L^\ast\right)_{l\beta}
    \left(D_L^\dagger\frac{Y_A}{\sqrt{2}}N^\ast\right)_{ki}
    \nonumber\\
    &\hspace{22mm}+\frac{g_2^2}{2}\frac{m_{d_k}m_{u_l}}{m_W^2}
    I_1(m_{d_k},m_{u_l})
    \left(V_{\mathrm{CKM}}^\ast\right)_{1k}
    \left(V_{\mathrm{CKM}}\right)_{l\alpha}
    \left(-U_L^\dagger 2(Y_B+Y_B^T)D_L^\ast\right)_{l\beta}
    \left(D_L^\dagger\frac{Y_A}{\sqrt{2}}N^\ast\right)_{ki}
    \nonumber\\
    &\hspace{22mm}+\frac{g_2^2}{2}\frac{m_{u_k} m_{e_l}}{m_W^2}I_1(m_{u_k},m_{e_l})
    \left(V_{\mathrm{CKM}}\right)_{k\beta}
    \left(U_{\mathrm{PMNS}}\right)_{li}
    \left(-U_L^\dagger 2(Y_B+Y_B^T)D_L^\ast\right)_{k\alpha}
    \left(U_L^\dagger \frac{-Y_A}{\sqrt{2}}E_L^\ast\right)_{1l}
    \nonumber\\
    &\hspace{22mm}-4g_2^2
    I_2(m_{u_k},m_{e_l})
    \left(V_{\mathrm{CKM}}\right)_{k\beta}
    \left(U_{\mathrm{PMNS}}\right)_{li}
    \left(-U_L^\dagger 2(Y_B+Y_B^T)D_L^\ast\right)_{k\alpha}
    \left(U_L^\dagger \frac{-Y_A}{\sqrt{2}}E_L^\ast\right)_{1l}
    ,\\
    &a(d_\alpha,d^c_\beta,\nu_i)=
    -\;g_2^2\frac{m_{u_k}m_{e_l}}{m_W^2}I_2(m_{u_k},m_{e_l})
    \left(V_{\mathrm{CKM}}\right)_{k\alpha}
    \left(U_{\mathrm{PMNS}}\right)_{li}
    \left(U_R^\dagger\frac{Y_A^\ast}{\sqrt{2}}D_R^\ast\right)_{k\beta}
    \left(U_R^\dagger 2(Y_B^\dagger+Y_B^\ast)E_R^\ast\right)_{1l}
    \nonumber\\
    &\hspace{22mm}-g_2^2 I_1(m_{u_k},m_{e_l})
    \left(V_{\mathrm{CKM}}\right)_{k\alpha}
    \left(U_{\mathrm{PMNS}}\right)_{li}
    \left(U_R^\dagger\frac{Y_A^\ast}{\sqrt{2}}D_R^\ast\right)_{k\beta}
    \left(U_R^\dagger 2(Y_B^\dagger+Y_B^\ast)E_R^\ast\right)_{1l}
    ,
\end{align}
\endgroup
where an implicit sum over repeated indices is meant and where we have used the following definition for the loop factors $I_1,I_2$~\cite{Passarino:1978jh,Denner:1991kt}:
\begin{align}
    &I_1(m_1,m_2)=-\frac{1}{16\pi^2}m_1m_2
    \left[ 
    \frac{m_1^2\log\left(\frac{m_1^2}{m_W^2}\right)}{(m_1^2-m_2^2)(m_1^2-m_{\phi_2}^2)(m_1^2-m_W^2)}
     +\frac{m_2^2\log\left(\frac{m_2^2}{m_W^2}\right)}{(m_2^2-m_1^2)(m_2^2-m_{\phi_2}^2)(m_2^2-m_W^2)}
    \right.\
    \nonumber\\
    &\hspace{7cm}\left.\
    +\;\frac{m_{\phi_2}^2\log\left(\frac{m_{\phi_2}^2}{m_W^2}\right)}{(m_{\phi_2}^2-m_1^2)(m_{\phi_2}^2-m_2^2)(m_{\phi_2}^2-m_W^2)}
    \right],
    \\
    &I_2(m_1,m_2)=-\frac{1}{16\pi^2}\left[
    \frac{m_1^4\log\left(\frac{m_1^2}{m_W^2}\right)}{(m_1^2-m_2^2)(m_1^2-m_{\phi_2}^2)(m_1^2-m_W^2)}
     +\frac{m_2^4\log\left(\frac{m_2^2}{m_W^2}\right)}{(m_2^2-m_1^2)(m_2^2-m_{\phi_2}^2)(m_2^2-m_W^2)}
    \right.\
    \nonumber\\
    &\hspace{7cm}\left.\
    +\;\frac{m_{\phi_2}^4\log\left(\frac{m_{\phi_2}^2}{m_W^2}\right)}{(m_{\phi_2}^2-m_1^2)(m_{\phi_2}^2-m_2^2)(m_{\phi_2}^2-m_W^2)}
    \right].
\end{align}
\twocolumngrid

\bibliographystyle{style}
\bibliography{references}
\end{document}